\documentclass[aps,twocolumn,showpacs]{revtex4}
\usepackage{amssymb,amsbsy,psfig,graphicx}

\def\be{\begin{equation}}
\def\ee{\end{equation}}
\def\bea{\begin{eqnarray}}
\def\eea{\end{eqnarray}}

\begin{document}

\title{
Structure of multiphoton quantum optics. I. Canonical \\
formalism and homodyne squeezed states}

\author{Fabio Dell'Anno}\email{dellanno@sa.infn.it}
\author{Silvio De Siena}\email{desiena@sa.infn.it}
\author{Fabrizio Illuminati}\email{illuminati@sa.infn.it}

\vspace{0.2cm}  \affiliation{\mbox{}Dipartimento di Fisica ``E. R.
Caianiello'', Universit\`a di Salerno, INFM UdR Salerno, INFN Sez.
Napoli -- G. C. Salerno, Via S. Allende, 84081 Baronissi (SA),
Italy}


\date{August 5, 2003}

\begin{abstract}
We introduce a formalism of nonlinear canonical transformations
for general systems of multiphoton quantum optics. For single-mode
systems the transformations depend on a tunable free parameter,
the homodyne local oscillator angle; for $n$-mode systems they
depend on $n$ heterodyne mixing angles. The canonical formalism
realizes nontrivial mixings of pairs of conjugate quadratures of
the electromagnetic field in terms of homodyne variables for
single--mode systems; and in terms of heterodyne variables for
multimode systems. In the first instance the transformations yield
nonquadratic model Hamiltonians of degenerate multiphoton
processes and define a class of non Gaussian, nonclassical
multiphoton states that exhibit properties of coherence and
squeezing. We show that such homodyne multiphoton squeezed states
are generated by unitary operators with a nonlinear
time--evolution that realizes the homodyne mixing of a pair of
conjugate quadratures. Tuning of the local--oscillator angle
allows to vary at will the statistical properties of such states.
We discuss the relevance of the formalism for the study of
degenerate (up-)down-conversion processes. In a companion paper,
{\bf ``Structure of multiphoton quantum optics. II. Bipartite
systems, physical processes, and heterodyne squeezed states''},
we provide the extension of the nonlinear canonical formalism to
multimode systems, we introduce the associated heterodyne multiphoton
squeezed states, and we discuss their possible experimental
realization.
\end{abstract}

\maketitle

\section{Introduction}
The study of the nonclassical states of light
has recently attracted renewed attention because
of the key role they may play, beyond the traditional realm
of quantum optics, in research fields of great current interest,
such as laser pulsed atoms and
molecules \cite{atom}; Bose--Einstein condensation
and atom lasers \cite{bec}; and quantum information
theory \cite{qci}.
Fundamental physical properties for
an efficient functioning of the quantum world such as
interferometric visibility, robustness of superpositions
against environmental perturbations, and degree of
entanglement are typically enhanced by exploiting states
which exhibit strong nonclassical features.

The simplest archetypal examples of nonclassical states are of
course number states, whose experimental realization is however
difficult to achieve. Moreover they share very few of the
coherence properties that would be desirable both in practical
implementations and in fundamental experiments. The most important
experimentally accessible nonclassical states are the
\emph{two--photon} squeezed states \cite{stoler,yuen,caves}. They
are Gaussian, exhibit several important coherence properties, and
can be obtained by suitably generalizing the notion of coherent
states \cite{glauber}. Squeezed states can be easily introduced by
linear Bogoliubov canonical transformations and the associated
eigenvalue equations for the transformed field operators
$b(a,a^{\dag})$: $b|\Psi>_{\beta}=\beta|\Psi>_{\beta}$. They can
be produced in the laboratory through the dynamical evolution of
parametric amplifiers \cite{parampl}, and provide a useful tool in
various areas of research. For instance, they have been proposed
to improve optical communications \cite{shapiro} and to measure
and detect weak forces and signals like gravitational waves
\cite{caves2}. Moreover, twin beams of bipartite systems, i.e.
two--mode squeezed states, are maximally entangled states, a
property of key importance in quantum computation and in quantum
information processing.

Realistic and scalable schemes of quantum devices and operations
with continuous variables might however require the realization of
multiphoton processes. In this respect, it is crucial to
investigate the existence and structure of \emph{multiphoton}
nonclassical states of light. A challenging goal is to define
suitable multiphoton generalizations of the effective Hamiltonian
description of two--photon down--conversion processes and
two--photon squeezed states. Natural candidates should be
nonclassical states obtained by nonlinear unitary evolutions
associated to anharmonic Hamiltonians and multiphoton
down--conversion processes. In turn, nonlinear unitary evolutions
might be of great importance in the implementation of universal
quantum computation with continuous variables (CV) systems
\cite{gottesmann,bartlett}. Experimental realizations could be
obtained by considering the dynamics of the polarizability in
nonlinear optical media \cite{bloember}: \be P_{i} = \epsilon_{0}
\left[ \chi^{(1)} E_{i} + \chi_{ijk}^{(2)} E_{j}E_{k} + ...
\chi_{i i_{1} i_{2} \cdot \cdot \cdot i_{n}}^{(n)} E_{i_{1}} \cdot
\cdot \cdot E_{i_{n}} + ... \right] \: , \label{Polzy} \ee where
$P$ is the polarization vector, $\chi^{(n)}$ the $n$--th order
susceptibility tensor, and $E$ the electric field. Implementation
of higher order multiphoton parametric processes involves several
terms in the expansion Eq.~(\ref{Polzy}). For instance,
$k$--photon parametric down conversions involve all contributions
at least up to the term with coupling $\chi^{(k)}$, whose strength
in nonlinear crystals is in general extremely weak for $k > 2$. It
must however be remarked that coherent atomic effects, such as
electromagnetically induced transparency and coherent population
trapping manipulations of photons in cavities provide new
promising techniques to generate large and lossless optical
nonlinearities \cite{radmatt}.

Phenomenological theories of multiphoton parametric amplification,
based on expansion Eq.~(\ref{Polzy}), were introduced by
Braunstein, Caves and McLachlan \cite{braunstein}, who considered
nonlinear interaction terms of the form $z_{k} a^{\dagger k} -
z_{k}^{*} a^k$ producing $k$--photon correlations. The problem was
numerically addressed by the authors, who showed that these
interactions generate squeezing and display remarkable
phase--space properties. Another interesting approach, but rather
abstract since it involves infinite powers of the canonical
creation and annihilation operators, was put forward in ref.
\cite{dariano} where generalized multiboson, non Gaussian squeezed
states were introduced. A crucial question left unanswered by the
above-mentioned attempts is that, although the first nonlinear
order in expansion Eq.~(\ref{Polzy}) can be associated, through
the linear Bogoliubov transformation, to an exactly diagonalizable
\emph{two--photon} Hamiltonian and to exact \emph{two--photon}
squeezed states \cite{yuen,caves}, higher order nonlinearities
have not been associated so far to exact Hamiltonian models of
multiphoton effective interactions in a simple and physically
transparent way.

In a series of recent papers \cite{noi1,noi2}, a first
step was realized in this direction by defining
canonical transformations that allow the exact
diagonalization of a restricted class of multiphoton
Hamiltonians.
In this formalism one adds to the linear
Bogoliubov transformation a nonlinear function
of a generic field quadrature.
The canonical conditions impose very stringent
relations on the parameters of the transformations,
and the resulting Hamiltonians describe a very peculiar
and restricted class of nonlinear interactions, not easily
amenable to realistic experimental realizations.

Studying the simplest case of a quadratic nonlinearity
\cite{noi1,noi2}, one determines multiphoton
squeezed states both in the case of nonlinear functions of the
first quadrature $X_{1} = (a+ a^{\dagger})/\sqrt{2}$ and in the
case of nonlinear functions of the second quadrature $X_{2} =- i
(a- a^{\dagger})/\sqrt{2}$. These states exhibit interesting nonclassical
statistics and squeezing in the
quadrature associated to the nonlinearity; they may be denoted as
\emph{single--quadrature multiphoton squeezed states} (SQMPSS).
The unitary operators associated to the
two transformations are a composition of squeezing, displacement,
and a nonlinear phase transformation \cite{noi2} (see also
\cite{wu}). The scheme, although limited to one--mode systems,
still provides some insights for experiments in quantum
information exploiting multiphoton processes \cite{weinfurter}. In
fact, the single--quadrature multiphoton squeezed states include
as a particular case the generalized ``cubic phase'' states
(originally introduced in the framework of quantum computation
\cite{gottesmann}) proposed by Bartlett and Sanders by adding
displacement and squeezing to the pure cubic phase transformation
\cite{bartlett}.
The single--quadrature canonical formalism
(and the associated single--mode, single--quadrature
multiphoton squeezed states) is thus very limited
because it amounts only to a
pure (nonlinear) phase transformations on a single
quadrature. Moreover, it does not allow nontrivial extensions
to multimode systems and nondegenerate processes.

In the present and in a companion paper, that
we shall denote as Part I and Part II,
we show that these
difficulties can be overcome and that it is
indeed possible to introduce a general
canonical formalism of multiphoton quantum optics.
In the present paper (Part I) we determine the most general
nonlinear canonical structure for single--mode systems
by introducing canonical transformations that depend on
generic nonlinear functions of homodyne combinations of
pairs of canonically conjugate quadratures.
The homodyne canonical formalism defines a class of
single--mode, homodyne multiphoton squeezed states;
it includes the single--mode, single--quadrature
multiphoton squeezed states as a particular case;
and introduces a tunable free parameter,
a local--oscillator mixing angle, which allows to interpolate
between different multiphoton model Hamiltonians and to
arbitrarily vary the field statistics of the states.
In the companion paper {\bf ``Structure of multiphoton quantum optics.
II. Bipartite systems, physical processes, and heterodyne squeezed
states''} (Part II) \cite{paper2} we extend
the multiphoton canonical formalism
to multimode systems. We show that such extension is
realized by nonlinear canonical transformations
of heterodyne combinations of field
quadratures. The scheme defines a structure of multimode, heterodyne
multiphoton squeezed states that reduce to the homodyne
states for single--mode systems. In Part II we also show
that the heterodyne squeezed states and the associated
effective multiphoton Hamiltonians can be
realized by relatively simple schemes of
multiphoton conversion processes \cite{paper2}.
In this way we introduce a complete hierarchy of canonical multiphoton
squeezed states: a) multimode heterodyne squeezed states; b) single--mode
homodyne squeezed states; and c) single--mode, single--quadrature
squeezed states.
In the limit of vanishing nonlinearity the multiphoton squeezed
states reduce to the standard (single-mode or multimode)
two--photon squeezed states.

To construct a general canonical formalism of
multiphoton quantum optics one must first circumvent
the restrictions following from the canonical conditions.
In particular, the prescription that
a general, nonlinear mode transformation be canonical
prevents the possibility of introducing
arbitrary nonlinear functions depending
simultaneously on two conjugate quadratures
$X_{1}$ {\it and} $X_{2}$.
In fact, if the form of the
nonlinear function is not constrained at all, the canonical conditions
force the nonlinear coupling to be trivially zero.

It is however possible to define a general canonical scheme by
introducing a simultaneous nonlinearity in two conjugate
quadratures if the nonlinear
part of the transformation is an arbitrary function
of the \emph{homodyne} combination
$\sqrt{2}|\eta| \left(
\cos{\theta} X_{1} + \sin{\theta} X_{2} \right)$
of the two quadratures
($\eta = |\eta|\exp{i\theta}$ arbitrary complex
number).
Such a canonical structure allows naturally for
a local oscillator angle $\theta$ mixing the
quadratures. The mixing is a physical process that
can be easily realized, e.g. by a beam splitter
positioned in front of a nonlinear crystal. Tuning
the continuous parameter $\theta$ allows then
to vary the physical properties, and in particular
the statistical properties, of the associated homodyne
multiphoton squeezed states.

The plan of the paper is as follows. In Section \ref{Formalism} we
develop the general formalism of nonlinear canonical transformations
for homodyne variables. In Section \ref{H&states} we study
the multiphoton Hamiltonians associated to
the nonlinear transformations, specializing to the case of quadratic
nonlinearity. We compute the wave functions of
coherent states associated to the transformations, as
functions of the mixing angle $\theta$. In Section
\ref{Uoperator} we determine the explicit form of the unitary
operators associated to the canonical transformations.
They are composed by the product of a
squeezing, a displacement, and a mixing operator with nonquadratic
exponent which combines conjugate quadratures.
In Section \ref{statistics}, we study the statistical properties
of the homodyne multiphoton squeezed states. We
compute the uncertainty products, and determine the condition for
"quasi-minimum" uncertainty.
We then study the
quasi--probability distributions, and the photon statistics.
We show that these properties depend strongly
on the local oscillator angle $\theta$. In Section
\ref{outlook}, we summarize our results and discuss the
outlook and extensions that are developed in the companion
paper Part II.

\section{Nonlinear canonical transformations
for homodyne variables}
\label{Formalism}
We could naively imagine
that the problem of introducing nonlinear canonical
transformations for a single bosonic mode $a$ should be solved by
adding to the standard Bogoliubov linear transformation an
arbitrary (sufficiently regular) Hermitian nonlinear function
$F(a, a^{\dagger})$ of the fundamental mode variables $a,
a^{\dagger}$:
\begin{equation}
b = \mu a + \nu a^{\dagger} + \gamma F(a, a^{\dagger}) \; .
\label{trasfimposs}
\end{equation}
Requiring that the transformed mode $b$ be bosonic, i.e. that
$[b,b^{\dagger}]=1$, and exploiting
the well known formulae
$$
[a, G(a, a^{\dagger})] =
\partial G(x,y)/\partial y|_{x=a, y=a^{\dagger}} \; ,
$$
$$
[a^{\dagger}, G(a,a^{\dagger})] = - \partial G(x, y)/\partial
x|_{x=a, y=a^{\dagger}} \; ,
$$
then the condition for the
transformation Eq. (\ref{trasfimposs}) to be canonical
reads
\bea 
&&|\mu|^{2} \, - \, |\nu|^{2} \, + 
\, |\gamma|^{2}[F,F^{\dag}] \, + \nonumber
\\ && \label{genccond} \, \\ 
&& \mu\gamma^{*}\frac{\partial F^{\dag}}{\partial
a^{\dag}}-\nu\gamma^{*}\frac{\partial F^{\dag}}{\partial a}
+\mu^{*}\gamma\frac{\partial F}{\partial
a}-\nu^{*}\gamma\frac{\partial F}{\partial a^{\dag}}=1 \,.
\nonumber 
\eea
It is very difficult to determine
the most general expression of the nonlinear function $F$, which
allows to satisfy the condition Eq.~(\ref{genccond}).
Nevertheless, if we assume that the nonlinear function
be hermitian, then canonical
generalizations of the Bogoliubov transformation do exist, and the
most general expression is in terms of arbitrary
Hermitian, nonlinear, analytic functions $F$ of homodyne linear
combinations of the fundamental mode variables. It reads: 
\be 
b =
\mu a + \nu a^{\dagger} + \gamma F(\eta^{*} a + \eta a^{\dagger})
\; , 
\label{boper2} 
\ee 
with $\eta \equiv | \eta | e^{i \theta }$
a complex number. Exploiting the functional dependence of $F$ on the
homodyne combination of the modes $a$ and $a^{\dagger}$, one finds
that the general relation Eq.~(\ref{genccond}) reduces to the following
algebraic constraints on the complex coefficients of the 
transformation: 
\bea
|\mu|^2 - |\nu|^2 & = & 1 \; , \nonumber \\
\nonumber \\
Re [e^{i \theta}(\mu \gamma^{*} - \nu^{*} \gamma)] & = & 0 \; .
\label{cacon2} 
\eea 
With the parametrization \be \mu = \cosh r \, , \, 
\nu = \sinh r e^{i\phi} \, , \, \gamma = |\gamma |
e^{i\delta} \; , 
\label{param} 
\ee 
we can express the canonical conditions Eqs.~(\ref{cacon2}) 
in the form of the transcendental equation 
\be 
\cosh r \cos(\theta-\delta)-\sinh r
\cos(\delta+\theta-\phi)=0 \; . 
\label{cacong} 
\ee 
Eq.~(\ref{cacong}) can be solved numerically. For instance, given some
fixed $r$, $\theta$ and $\phi$ we can find numerical solutions for
the phase variable $\delta$, which can be used as an adjustable
parameter to assure the canonical structure of the transformation. 
Alternatively, we can look for particular analytical solutions of 
Eq.~(\ref{cacong}): letting $\phi=0$, we obtain the simplified expression 
\be 
\tan \theta \tan \delta =-e^{-2r} \; .
\label{Ccon}
\ee 
For given values of the local oscillator angle $\theta$
this is a relation between the phase $\delta$ of the nonlinearity
and the strength $r$ of the squeezing; e.g., fixing
$\theta=\pm\pi/4$, we get $\tan \delta=\mp e^{-2r}$. Setting
$\theta=-\delta$ implies instead $\tan\delta=e^{-r}$. Of course,
Eq.~(\ref{cacong}) admits infinite solutions which correspond to a
great variety of nonlinear canonical operators. We can however
select more stringent conditions, imposing 
\be 
\delta - \theta =
\pm \frac{\pi}{2} \pm k \pi \, \, , \, \, \delta + \theta - \phi =
\pm \frac{\pi}{2} \pm h \pi \, , 
\label{cacons} 
\ee 
with $k, h$ arbitrary integers. 
This choice allows to satisfy Eq.~(\ref{cacong})
at the price of eliminating one degree of freedom
from the problem. In conditions Eqs.~(\ref{cacons}) it is
obviously sufficient to consider $k = h = 0$. 
From Eqs.~(\ref{cacon2}) and (\ref{cacong}) it is evident 
that the modulus of $\eta$ is irrelevant in the determination 
of the canonical constraints of the transformations. 
Therefore, from now on we set
$|\eta| = 1/\sqrt{2}$. In this way, the homodyne character of the
transformation scheme becomes fully evident. We can in fact
express the transformation Eq.~(\ref{boper2}) in terms of the
rotated homodyne quadratures $X_{\theta}, P_{\theta}$ defined as
\bea 
X_{\theta} & = & \left( ae^{- i\theta} +
a^{\dagger}e^{i\theta}\right) /\sqrt{2}
= X_{1} \cos \theta + X_{2} \sin \theta \, \, , \nonumber \\
&& \nonumber \\
P_{\theta} & \equiv & X_{\theta + \pi/2} = - X_{1} \sin \theta +
X_{2} \cos \theta \; , 
\label{defxp} 
\eea 
with $[X_{\theta}, P_{\theta}] = i$. The transformed mode $b$ 
can then be expressed in terms of the rotated mode 
$a_{\theta } = ae^{- i\theta}$, or of
the rotated quadrature $X_{\theta}$, as 
\be 
b = \tilde{\mu}
a_{\theta} + \tilde{\nu} a_{\theta}^{\dagger} + \gamma
F(X_{\theta}) \; , 
\label{boperg} 
\ee 
with the rotated parameters
\be 
\tilde{\mu} = \mu e^{i \theta} \; ; \; \; \; \; \tilde{\nu} =
\nu e^{-i\theta} \; . 
\label{munup} 
\ee 
From Eqs.~(\ref{defxp}),~(\ref{boperg}),~(\ref{munup}) 
the canonical conditions Eqs.~(\ref{cacon2}) can be expressed 
in terms of the rotated parameters as 
\bea
|\tilde{\mu}|^{2} - |\tilde{\nu}|^2 & = & 1 \; , \nonumber \\
&& \nonumber \\
Re [(\tilde{\mu} \gamma^{*} - {\tilde{\nu}}^{*} \gamma)] & = & 0
\; . 
\label{cancon} 
\eea 
The form Eq.~(\ref{boperg}) of the
canonical transformation Eq.~(\ref{boper2}) allows the
straightforward determination of the coherent states of the
transformed modes, and of the unitary operators associated to the
transformations, as we will show in the following Sections.
Obviously, when $\gamma = 0$ one recovers the standard linear
Bogoliubov transformations and the structure of standard
two--photon squeezed states.

\section{Multiphoton Hamiltonians and multiphoton squeezed states}

\label{H&states} 

We now consider the Hamiltonians that can be
associated to the canonical transformations Eq.~(\ref{boper2}). 
We will restrict ourselves to the case of quadratic Hamiltonians
diagonal in the transformed modes: 
\be 
H = b^{\dagger} b \; , 
\ee
whose general expression in terms of $a, a^{\dagger}$ reads
\begin{eqnarray}
H & = & |\nu|^{2} + \left( |\mu|^{2} + |\nu|^{2} \right)
a^{\dag}a + \left( \mu \nu^{*}a^{2} + h. c. \right)
\nonumber \\
&& \nonumber \\
& + & \left( \mu^{*}\gamma a^{\dag}F
+ \nu^{*}\gamma a F + |\gamma|^{2}F^{\dag}F
+ h. c. \right) \; .
\label{ham}
\end{eqnarray}
In terms of the rotated quadratures $X_{\theta}, P_{\theta}$,
and exploiting the canonical constraints Eq.~(\ref{cacong}),
the Hamiltonian Eq.~(\ref{ham}) reads
\bea
H & = & -\frac{1}{2} + \frac{1}{2}|\tilde{\mu} + \tilde{\nu}|^{2}
X_{\theta}^{2}
+ \frac{1}{2}|\tilde{\mu} - \tilde{\nu}|^{2}
P_{\theta}^{2} \nonumber \\
&& \nonumber \\
& + & |\gamma|^{2}F^{2}(X_{\theta}) +
\frac{i}{\sqrt{2}}\gamma^{*}(\tilde{\mu} -
\tilde{\nu})\{P_{\theta},F(X_{\theta})\} \, , 
\label{hamxp} 
\eea
where $\{\cdot,\cdot\}$ denotes the anticommutator. The
Hamiltonian Eq.~(\ref{hamxp}) is further simplified 
by exploiting the conditions Eqs.~(\ref{cacons}):
\begin{equation}
H=\frac{e^{\pm 2r}}{2}X_{\theta}^{2}-\frac{1}{2}+\frac{e^{\mp
2r}}{2}[P_{\theta}\pm\sqrt{2}|\gamma|e^{\pm r}F(X_{\theta})]^{2}
\; . \label{hamxp2}
\end{equation}
Eq.~(\ref{hamxp2}) is of the same form obtained in 
Refs.~\cite{noi1}--\cite{noi2}, but with the fundamental
difference that the nonlinearity is now placed on the
homodyne-mixed, rotated quadrature $X_{\theta}$ rather
than on a single one of the original $X_{i}'s$.
Moreover the mixing depends on a tunable free parameter,
the local oscillator angle $\theta$.
Eq.~(\ref{hamxp2}) shows that the variable $X_{\theta}$ is
squeezed and that its conjugate variable, in the sense of being
antisqueezed of a corresponding amount, is the
``generalized momentum'' $P_{\theta} \pm\sqrt{2}
|\gamma| e^{\pm r} F (X_{\theta})$. The associated
quasi--probability distributions are then squeezed along a rotated
axis, as will be shown in the following Sections.

\subsection{The case of quadratic nonlinearity}

In studying the statistical properties of the coherent states
associated to the Hamiltonians Eqs.~(\ref{ham})--(\ref{hamxp}) we
will specialize to the case of the lowest possible nonlinearity in
powers of the homodyne rotated quadratures, i.e. we will consider
the quadratic form 
\be
F (X_{\theta}) = X_{\theta}^2 \; .
\label{quadratic}
\ee
Inserting Eq.~(\ref{quadratic}) in Eqs.~(\ref{ham})-(\ref{hamxp}),
the corresponding four-photon Hamiltonian reads
\begin{eqnarray}
H_{4p} & = & A_{0} + \left( A_{1}a^{\dag}
+ A_{2}a^{\dag2} + A_{3}a^{\dag3}
+ A_{4}a^{\dag4} + h. c. \right) \nonumber \\
&& \nonumber \\
& + & B_{0}a^{\dag}a
+ B_{1}a^{\dag2}a^{2} \nonumber \\
&& \nonumber \\
& + &
Ca^{\dag2}a + D a^{\dag3}a + h. c. \; ,
\label{ham4ph}
\end{eqnarray}
where the coefficients $A_{i}, B_{i}, C, D$ are
\begin{eqnarray}
&& A_{0}=|\nu|^{2}+\frac{3}{4}|\gamma|^{2} \; , \;
A_{1}=\frac{1}{2}\mu^{*}\gamma+\frac{3}{2}\nu\gamma^{*}+
e^{2i\theta}\nu^{*}\gamma \; , \nonumber \\
&& \nonumber \\ 
&& A_{2}=\frac{3}{2}e^{2i\theta}|\gamma|^{2}+\mu^{*}\nu \; ,
\; A_{3}=\frac{1}{2}e^{2i\theta}\mu^{*}\gamma
+\frac{1}{2}e^{2i\theta}\nu\gamma^{*} \, , \nonumber \\
&& \nonumber \\ 
&& A_{4}=\frac{1}{2}e^{4i\theta}|\gamma|^{2} \; ,
\; \ B_{0}=|\mu|^{2}+|\nu|^{2}+3|\gamma|^{2} \; , \nonumber \\
&& \nonumber \\ 
&& B_{1}=\frac{3}{2}|\gamma|^{2}\; , \; C =
\frac{1}{2}e^{2i\theta}(\mu\gamma^{*}+\nu^{*}\gamma)
+\mu^{*}\gamma+\nu\gamma^{*} \; , \nonumber \\
&& \nonumber \\ 
&& D=e^{2i\theta}|\gamma|^{2} \; .
\label{coefficients}
\end{eqnarray}
The Hamiltonian Eq.~(\ref{ham4ph}) describes one--, two--, three--
and four--photon processes with effective linear and nonlinear
photon--photon interactions associated to degenerate parametric
down conversion processes in nonlinear media.
We see that the Hamiltonian coefficients Eq.~(\ref{coefficients})
crucially depend on the homodyne angle $\theta$, allowing for a
great freedom in searching for physical implementations of
multiphoton states by processes associated to higher order
nonlinear susceptibilities.

\subsection{Homodyne multiphoton squeezed states}

We define the coherent states
$|\Psi\rangle_{\beta}$ associated to the Hamiltonian
Eq.~(\ref{ham}) as the eigenstates (with complex
eigenvalue $\beta = |\beta|e^{i\xi}$) 
of the transformed annihilation operator $b$:
\be
b  |\Psi\rangle_{\beta} \, = \, \beta |\Psi\rangle_{\beta} \, .
\label{eigenvalue}
\ee
Choosing the representation $\Psi_{\beta}(x_{\theta}) \equiv
\langle x_{\theta}|\Psi\rangle_{\beta}$ in which the
homodyne rotated quadrature $X_{\theta}$ is diagonal,
the eigenvalue equation Eq.~(\ref{eigenvalue}) reads
\begin{eqnarray}
\partial_{x_{\theta}}\Psi_{\beta}(x_{\theta})
& = & -\frac{1}{\tilde{\mu} - \tilde{\nu}} \big[
\left( \tilde{\mu} + \tilde{\nu} \right) x_{\theta}
\nonumber \\
& + & \sqrt{2} \gamma F(x_{\theta})
- \sqrt{2}\beta \big] \Psi_{\beta}(x_{\theta}) \; ,
\label{diffeq}
\end{eqnarray}
where we have used $P_{\theta} = - i \partial_{x_{\theta}}$. The
general solution of Eq.~(\ref{diffeq}) is
\be
\Psi_{\beta}(x_{\theta}) = {\cal{N}} \exp \left[ -\frac{a}{2}
x_{\theta}^{2} + cx_{\theta} - b
\int^{x_{\theta}}dyF(y) \right] \, ,
\label{wf}
\ee
where
$$
{\cal{N}} =
\left(
\frac{\pi}{Re[a]}
\right)^{-1/4} \exp \left[
-\frac{(Re[c])^{2}}{2Re[a]} \right]
$$
is the normalization, and the coefficients read
\be
a=\frac{\tilde{\mu} + \tilde{\nu}}{\tilde{\mu}
- \tilde{\nu}} \, , \quad
b=\frac{\sqrt{2}\gamma}{\tilde{\mu} -
\tilde{\nu}} \, , \quad
c=\frac{\sqrt{2}\beta}{\tilde{\mu} -
\tilde{\nu}} \; .
\label{coef2}
\ee
It can be easily verified that it is always
$Re[a]>0$ and $Re[b]=0$. The wave function 
can then be expressed in the form
$$
\Psi_{\beta}(x_{\theta}) = \left[\frac{Re[a]}{\pi}\right]^{1/4}
\exp\left[-\frac{Re[a]}{2}\left(x_{\theta}-\frac{Re[c]}{Re[a]}
\right)^{2}\right] \times
$$
\be
\exp\left[-i\left(Im[b]\int^{x_{\theta}}dyF(y)
+\frac{Im[a]}{2}x_{\theta}^{2}-Im[c]x_{\theta}\right)\right] \, .
\label{wf2}
\ee
By recalling the definition of the parameters
Eqs.~(\ref{param}) and the canonical conditions
Eqs.~(\ref{cacons}), Eq.~(\ref{wf2}) reduces to
$$
\Psi_{\beta}(x_{\theta}) =
A \exp \left\{
- \frac{e^{\pm 2r}}{2} \left[
x_{\theta} - \sqrt{2} |\beta| e^{\mp r}
\cos(\xi-\theta) \right]^{2} \right\} \times
$$
\be
\exp \left\{
i\sqrt{2}e^{\pm r}
\left[ |\beta| \sin (\xi - \theta)x_{\theta}
- |\gamma| \int^{x_{\theta}}dy
F(y) \right] \right\} \; ,
\label{wf3}
\end{equation}
where $A = \pi^{-1/4}e^{\pm r/2}$.
The wave function Eq.~(\ref{wf3}) has
a Gaussian density in the variable $x_{\theta}$,
with a squeezed variance, while
both the phase and the density depend crucially
on the homodyning, i.e. on the local oscillator
angle $\theta$. We name therefore the states
Eqs.~(\ref{eigenvalue})--(\ref{wf3}) Homodyne
Multiphoton Squeezed States (HOMPSS).

The dependence on the homodyning $\theta$ has
important physical implications, especially
with regard to the statistical properties
of the HOMPSS. In general the
states Eqs.~(\ref{wf2})--(\ref{wf3}) in
the $X_\theta$-diagonal representation display non
Gaussian terms in the phase, such as
a cubic phase term in $X_{\theta}$ in the case of
a quadratic nonlinearity. The nontrivial
structure of the HOMPSS emerges clearly
when writing in terms of the
original quadratures, for instance in the
$X_{1}$-diagonal representation
$\Psi_{\beta} (x)
\equiv \langle x | \Psi \rangle_{\beta}$.
Specializing to the quadratic nonlinearity
$F (X_{\theta}) = X_{\theta}^2$ and 
for a generic angle $\theta \neq 0,\pi$, one has
\begin{equation}
\Psi_{\beta}(x) =
{\cal{N}}
\exp \left\{
ia x^{2} + b x \right\}
Ai \left[
\frac{c x+d}{c^{2/3}} \right] \; ,
\label{airy}
\end{equation}
where $Ai [\cdot ]$ denotes the Airy function, and the
coefficients $a, b, c, d$ are given by
\begin{eqnarray}
a & = & -\frac{1}{2}\cot\theta \; \; \;  , \; \; \; \;
b = \frac{\mu-\nu}{2\sqrt{2}\gamma \sin^{2}\theta} \; ,
\nonumber \\ \ && \nonumber \\
c & = & \frac{-i(\tilde{\mu} - \tilde{\nu})}{\sqrt{2}
\gamma \sin^{3}\theta} \; \; , \; \; \;
d = \frac{(\mu-\nu)^{2}}{8\gamma^{2}\sin^{4}\theta}
-\frac{\beta}{\gamma \sin^{2}\theta}
\; . \nonumber
\end{eqnarray}
The general non Gaussian character of the
HOMPSS is thus apparent when writing them
in the original field--quadrature representation.

\subsection{Reduction to single--quadrature
multiphoton squeezed states}

When considering the special cases
$\theta = 0$ and $\theta = \pi/2$
the HOMPSS reduce to the Single--Quadrature
Multiphoton Squeezed States (SQMPSS) previously
introduced in Refs.~\cite{noi1,noi2}.
In these two special cases the canonical
transformations Eq.~(\ref{boperg}) reduce to
\be b = \mu a + \nu a^{\dagger} + \gamma
F(X_{i}) \, \, , \, \, i = 1, 2 \; \; ,
\label{boper}
\ee
where $i=1$ for $\theta = 0$ and
$i=2$ for $\theta = \pi/2$.

The associated coherent states are defined as the eigenstates of
$b$ with eigenvalue $\beta$; in the case
of zero phase difference between $\mu$ and $\nu$, and
parameterizing $\beta$ in terms of the coherent amplitude
$\alpha$, $\beta = \mu \alpha + \nu \alpha^{*}$ ($\alpha =
\alpha_{1} + i \alpha_{2}$), they read \cite{noi1,noi2}:
$$
\Psi_{\beta}^{\gamma F} (x_{i}) = \frac{1}{\sqrt{\pi e^{- 2
r_{i}}}} \exp\{- \frac{(x_{i} - x_{i}^{(0)})^2}{2 e^{- 2 r_{i}}}\}
\times
$$
\be
\exp\{i[c_{i} x_{i} + e^{r_{i}} {\tilde \gamma}_{i} G(x_{i})]\} \;
, \; \; i = 1, 2 \; ,
\label{wavfun}
\ee
where
\bea
&& r_{1} = r \; , \; r_{2} = - r \; , \nonumber \\
&& {\tilde \gamma}_{1} = Im (\gamma) \; ,
{\tilde \gamma}_{2} = - Re (\gamma) \; , \nonumber \\
&& G(z) = \int_{0}^{z} F(y) dy \, , \nonumber \\
&& x_{1}^{(0)} = \sqrt{2} \alpha_{1} \; , \;
x_{2}^{(0)} = - \sqrt{2} \alpha_{2} \; , \nonumber \\
&& c_{1} = \sqrt{2} \alpha_{2} \; , \; c_{2} = - \sqrt{2}
\alpha_{1} \; . \nonumber  \;
\eea
The expression Eq.~(\ref{wavfun}) for the SQMPSS
shows that the quadrature associated to the nonlinearity is squeezed, 
while, in the chosen
representation, the nonlinear function $F$ enters in the phase of
the wave packet. Explicit non Gaussian densities are again realized
if we adopt the ``coordinate'' representation (in which $X_{1}$ is
diagonal) when the nonlinearity is placed on $X_{2}$, or viceversa.

\section{Unitary operators} 

\label{Uoperator}

The expression Eq.~(\ref{wf3}) allows to identify, in terms of
the homodyne rotated quadratures, the unitary operator $U_{hom}$
associated to the canonical transformation Eq.~(\ref{boperg}),
such that the HOMPSS are obtained by applying $U_{hom}$ to the vacuum:
\begin{equation}
|\Psi \rangle_{\beta} =
U_{\hom} |0 \rangle \; ,
\end{equation}
where $|0 \rangle$ is the
vacuum state of the original field mode $a$: $a|0 \rangle = 0$.
The unitary operator $U_{\hom}$ then reads
\begin{equation}
U_{\hom} = U_{\theta}(X_{\theta})
D_{\theta}(\alpha_{\theta})S_{\theta}(\zeta_{\theta}) \; ,
\end{equation}
where
$$
D_{\theta}(\alpha_{\theta}) = \exp \left( \alpha_{\theta}
a^{\dag}_{\theta} - \alpha^{*}_{\theta} a_{\theta} \right) \; ,
$$
is the standard Glauber displacement operator with
$\alpha_{\theta} = {\tilde{\mu}}^{*} \beta
- \tilde{\nu} \beta^{*}$, and
$$
S_{\theta}(\zeta) =
\exp \left(-\frac{\zeta_{\theta}}{2}a^{\dag2}_{\theta} +
\frac{\zeta_{\theta}^{*}}{2}a^{2}_{\theta} \right) \; ,
$$
is the standard squeezing operator with $\zeta_{\theta} = r
e^{i(\phi-2\theta)}$. Finally, the ``mixing'' operator
$U_{\theta}$ reads
\be
U_{\theta}(X_{\theta}) = \exp \left[ -i
Im[b] \int^{X_{\theta}} dY F(Y) \right] \; ,
\label{Uphase}
\ee
where, exploiting the canonical conditions Eqs.~(\ref{cacons}),
$Im[b]=\pm \sqrt{2}|\gamma|e^{\pm r}$, and the abstract
operatorial integration acquires a precise operational meaning in
any chosen representation.
If we specify to the case of quadratic nonlinearity $F(X_{\theta})
= X_{\theta}^{2}$, we have:
\bea
U_{\theta}(X_{\theta}) & = & \exp
\left[ -iIm[b]3^{-1} 2^{-3/2}
(a e^{- i\theta} + a^{\dagger}e^{i\theta})^3 \right] \nonumber \\
& = & \exp \left[ -i Im[b]3^{-1}2^{-3/2} (a^3 e^{-3i\theta} +
a^{\dagger 3}e^{3i\theta} \right. \nonumber \\
&+&\left. 3 a^{\dagger} a^2 e^{- i\theta} + 3 a^{\dagger 2} a
e^{i\theta} + 3 a e^{- i\theta} + 3 a^{\dagger} e^{i\theta})
\right] \; . \nonumber \\ 
&&
\eea
The structure of $U_{\theta}$ is elucidated by expressing it in
terms of the field quadratures:
\bea
U_{\theta}(X_{\theta}) & = &
\exp \left[ -i Im[b]3^{-1} \left( X_{1}^3 \cos^3 \theta
+ X_{2}^3 \sin^3 \theta \right. \right. \nonumber \\
& + & \left. \left. 3 X_{1}^2 X_{2} \cos^2 \theta \sin \theta + 3
X_{1} X_{2}^{2} \sin^2 \theta \cos \theta \right. \right.
\nonumber \\
& + & \left. \left. 3 i X_{1} \cos^2 \theta \sin \theta + 3 i
X_{2} \sin^2 \theta \cos \theta ) \right) \right] \; . \nonumber \\ 
&&
\label{Uph3} 
\eea
The unitary operator $U_{\theta}$
Eq.~(\ref{Uph3}) depends on all powers of the conjugate field
quadratures up to $n+1$ if the nonlinearity $F$ is a power of
order $n$ of the homodyne quadrature $X_{\theta}$.
$U_{\theta}$ is a mixing operator: it mixes
nontrivially the original quadratures, depending on the values of
the local oscillator angle, except for the special cases $\theta
=0$ and $\theta = \pi/2$ treated in Refs.~\cite{noi1,noi2}, where
the mixing disappears and the nonlinearity is a simple power of a
single field quadrature (either $X_{1}$ or $X_{2}$). Obviously,
the quadratic nonlinearity is only one of a large class of
possible choices allowed for $F$ in Eq.~(\ref{Uphase}), so that
many complex nonlinear unitary homodyne mixing of the original
field quadratures can be realized.

In the special cases $\theta =0$ and
$\theta = \pi/2$ the HOMPSS reduce
to the SQMPSS. Then, from the choice of representation
of Eq.~(\ref{wavfun}) one determines
the form of the unitary operators that produce
these particular multiphoton squeezed states:
\be
U_{i} = exp [i e^{r_{i}} {\tilde
\gamma}_{i} G(X_{i})] D(\alpha)S(r), \label{uniop} \;
\ee
where $S(r)$ is the one--mode squeezing operator for $\phi = 0$ and
$D(\alpha)$ is the Glauber displacement operator.
We see that in these particular cases the
nonlinear part of the transformation adds to squeezing
and displacement a pure nonlinear phase term in one of
the quadrature. We also notice that with the choice
$F = X_{1}^2$ we recover the cubic phase states
proposed in Ref.~\cite{bartlett} as a possible tool
for the realization of quantum logical gates.

\section{Statistical properties and homodyne angle tuning}

\label{statistics}

\subsection{Uncertainty products}

When considering the statistical properties of the HOMPSS we first
study the behavior of the uncertainties in the homodyne
quadratures $X_{\theta}, P_{\theta}$. Let us express the
generalized variables in terms of the transformed mode operators
$b$ and $b^{\dagger}$:
\begin{equation}
X_{\theta}=\frac{1}{\sqrt{2}}
[({\tilde{\mu}}^{*} - {\tilde{\nu}}^{*})b
+ (\tilde{\mu} - \tilde{\nu}) b^{\dag}] \; ,
\end{equation}
\be  P_{\theta}=\frac{i}{\sqrt{2}}
[(\tilde{\mu} + \tilde{\nu})b^{\dag}
-({\tilde{\mu}}^{*} + {\tilde{\nu}}^{*})b
- 2iIm(\tilde{\mu}\gamma^{*}
- \tilde{\nu}^{*} \gamma)F(X_{\theta})] \; .
\ee
The above lead to the following expressions for the uncertainties:
\begin{eqnarray}
\Delta^{2}X_{\theta} & = &
\frac{1}{2}|\tilde{\mu} - \tilde{\nu}|^{2} \; ,
\nonumber \\
\Delta^{2}P_{\theta} & = &
\frac{1}{2}|\tilde{\mu} + \tilde{\nu}|^{2}
+ 2Im^{2}[{\tilde{\mu}}^{*}\gamma -
\tilde{\nu}\gamma^{*}] \times
\nonumber \\
&& (\langle F^{2}\rangle_{\beta}
- \langle
F \rangle_{\beta}^{2})-2Im[{\tilde{\mu}}^{*}\gamma
- \tilde{\nu} \gamma^{*}] \times
\nonumber \\
&& Im[(\tilde{\mu} + \tilde{\nu})
\langle [F,b^{\dag}] \rangle_{\beta}] \; ,
\label{dp2}
\end{eqnarray}
where $\langle \cdot \rangle_{\beta}$ denotes the expectation
value in the HOMPSS $|\Psi \rangle_{\beta}$, and $[\cdot ,\cdot ]$ 
denotes the commutator. It is evident that the nonlinearity
affects only $ \Delta P_{\theta}$, as the second of Eqs.~(\ref{dp2})
explicitly depends from the form of the function $F$.
Considering the quadratic form for $F$ and assuming the canonical
conditions Eqs.~(\ref{cacons}), Eqs.~(\ref{dp2}) become : \bea
\Delta^{2}X_{\theta} & = & \frac{1}{2}e^{\mp 2r} \; ,
\nonumber \\
\Delta^{2}P_{\theta} & = & \frac{1}{2}e^{\pm 2r}
\nonumber \\
& + & e^{\mp 2r}|\gamma|^{2}\{ 1 + 4|\beta|^{2}
+ 4Re[e^{-2i\theta}\beta^{2}] \} \; .
\label{dxps}
\eea
If we now consider the uncertainty product, and if we define
$\beta=|\beta|e^{i\xi}$, Eqs.~(\ref{dxps}) yield
\begin{equation}
\Delta^{2}X_{\theta}\Delta^{2}P_{\theta}
=\frac{1}{4}+\frac{1}{2}|\gamma|^{2}e^{\mp
4r}\{1+4|\beta|^{2}+4|\beta|^{2}\cos 2(\xi-\theta)\} \; .
\label{dxps1}
\end{equation}
It is to be remarked that this last relation attains its
minimum for $\xi-\theta=\pm\frac{\pi}{2}$:
\begin{equation}
\Delta^{2}X_{\theta}\Delta^{2}P_{\theta}
=\frac{1}{4}+\frac{1}{2}|\gamma|^{2}e^{\mp 4r} \; . 
\label{dxps2}
\end{equation}
Eq.~(\ref{dxps2}) can be seen as a "quasi-minimum" uncertainty
relation; in fact, although the second term is not exactly zero,
for small nonlinearities it will be surely very small with respect
to the first term (i.e. the Heisenberg minimum), due both
to $|\gamma| < 1$ and to the decreasing contribution of
the exponential for a suitable choice of the sign of $r$.

\subsection{Average photon number}

\begin{figure}
\begin{center}
\includegraphics*[width=8cm]{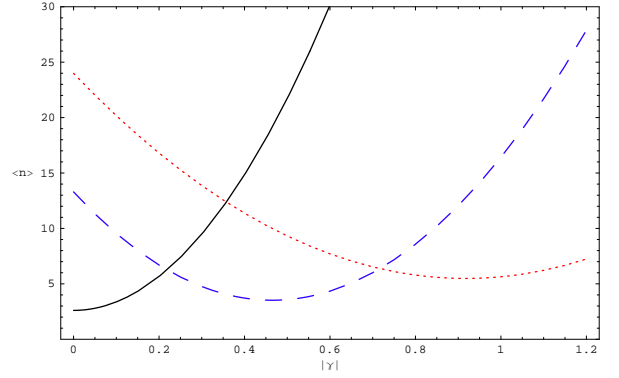}
\end{center}
\caption{Mean photon number $\langle n \rangle$ as function of
$|\gamma|$, for a HOMPSS with magnitude of squeezing $r = 0.8$,
coherent squeezed amplitude $\beta = 3$, and different mixing
angles: $\theta = 0$ (full line); $\theta = \frac{\pi}{6}$ (dashed
line); and $\theta = \frac{\pi}{4}$ (dotted line).}
\label{meannumb1}
\end{figure}

\begin{figure}
\begin{center}
\includegraphics*[width=8cm]{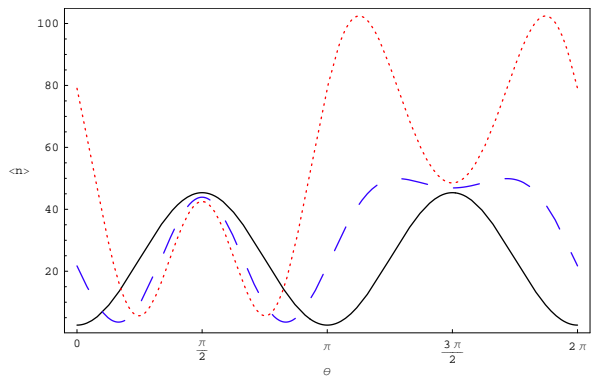}
\end{center}
\caption{The mean photon number
$\langle n \rangle$ as a function of $\theta$, for a
HOMPSS with $r=0.8$, $\beta=3$ and different
strengths of the nonlinearity:
$|\gamma| = 0$ (full line);
$|\gamma| = 0.5$ (dashed line);
and $|\gamma| = 1$ (dotted line).}
\label{meannumb2}
\end{figure}

We now turn to the calculation of the average number of photons
$\langle n \rangle = \langle a^{\dag}a \rangle$ in a HOMPSS
$|\Psi\rangle_{\beta}$. We specialize to the case of a quadratic
nonlinearity. We will show that $\langle n \rangle$ is strongly
affected both by the strength $|\gamma|$ of the nonlinearity and
by the mixing angle $\theta$. In Fig.~\ref{meannumb1} we study the
behavior of $\langle n \rangle$ as a function of $|\gamma|$ for
fixed values of the squeezed coherent amplitude $\beta$ and of the
magnitude $r$ of the squeezing, and for three different values of
the mixing angle $\theta$.
Due to the
canonical condition $\phi = 2 \theta$, there cannot
be pure squeezing for $\theta \neq 0$ even in absence
of the nonlinearity, and thus at $\gamma = 0$
we have different initial average numbers of photons
depending on the value of $\theta$.
The analytic expression for $\langle
n\rangle $ for a homodyne transformation with
$|\gamma|=0$ and $\phi = 2\theta$ reads:
\begin{equation}
\langle n \rangle = |\beta|^{2}\cosh 2r
-Re[\beta^{*2}e^{2\imath\theta}]\sinh 2r+\sinh^{2}r \; .
\end{equation}
We see from Fig.~\ref{meannumb1} that $\langle n \rangle$ needs
not show a monotonic behavior as a function of the nonlinearity.
It is in fact very sensitive to the mixing angle $\theta$, and
although it eventually always grows for sufficiently large values
of $|\gamma|$, it can however initially decrease, depending on the
mixing angle $\theta$, and then increasing again very slowly at
larger values of $|\gamma|$.
This behavior suggests that $\langle n \rangle$ will show even
more remarkable properties when varying $\theta$ for different,
fixed values of $|\gamma|$. In Fig.~\ref{meannumb2} we analyze the
behavior of $\langle n\rangle $ as function of $\theta$, at fixed
$|\gamma|$.
For $|\gamma| = 0$ we recover the oscillatory behavior of $\langle
n \rangle$ as a function of the squeezing phase $\phi = 2 \theta$
of the standard two--photon squeezed state. When letting $\theta$
vary at fixed finite values of $|\gamma|$, the oscillations of
$\langle n \rangle$ become faster, and both peaks and bottoms
quickly rise to very large values. Therefore the nonlinearity
plays the role of a ``quantum coherent pump'' allowing for very
large average numbers of photons even in the case of lowest,
quadratic nonlinearity.

\subsection{Quasi--probability distributions and
phase--space analysis}

We now turn to the study of the statistics of direct, heterodyne,
and homodyne detection for the homodyne multiphoton squeezed
states $|\Psi\rangle_{\beta}$, showing in particular how the
statistics is significantly modified by the tuning of the mixing
angle $\theta$. We will consider the HOMPSS Eq.~(\ref{wf3}),
corresponding to the canonical conditions Eqs.~(\ref{cacons}), and
we will specialize to the lowest nonlinear function $F(x_{\theta})
= x_{\theta}^2$.

\begin{figure}
\begin{center}
\includegraphics*[width=8.2cm]{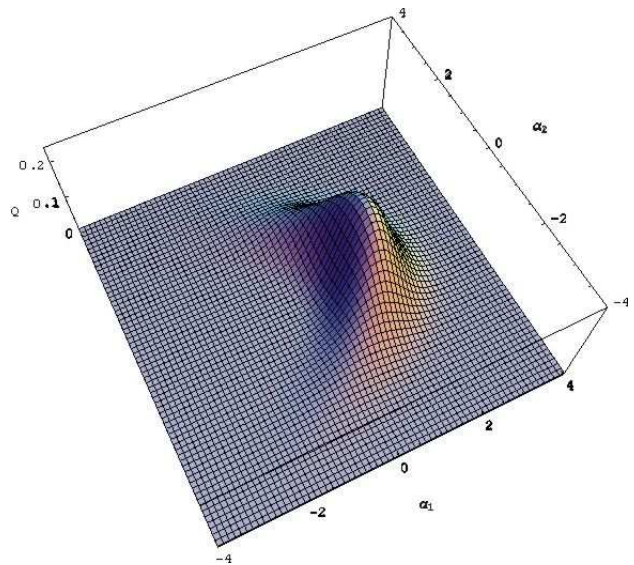}
\end{center}
\caption{Plot of the $Q-$function, with $r=0.8$, $|\gamma|=0.4$,
$\beta=3$, $\theta=\frac{\pi}{2}$, for the canonical conditions
$\delta-\theta=-\frac{\pi}{2}$,
$\delta+\theta-\phi=\frac{\pi}{2}$.}
\label{Q1}
\end{figure}

\begin{figure}
\begin{center}
\includegraphics*[width=8.2cm]{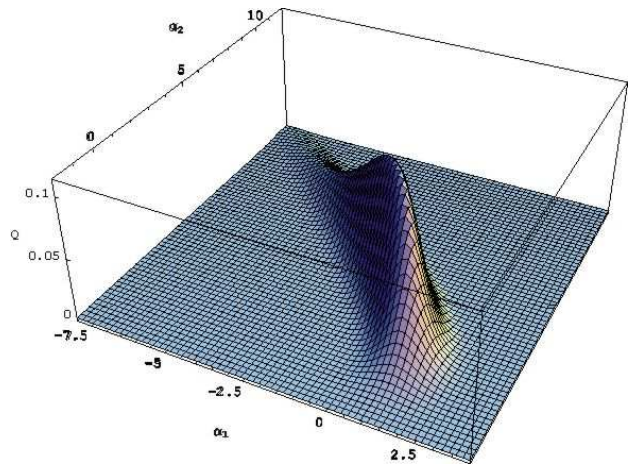}
\end{center}
\caption{$Q-$function, with $r=0.8$, $|\gamma|=0.4$, $\beta=3$,
$\theta=\frac{\pi}{3}$, for the canonical conditions
$\delta-\theta=-\frac{\pi}{2}$,
$\delta+\theta-\phi=\frac{\pi}{2}$.}
\label{Q2}
\end{figure}

\begin{figure}
\begin{center}
\includegraphics*[width=8.2cm]{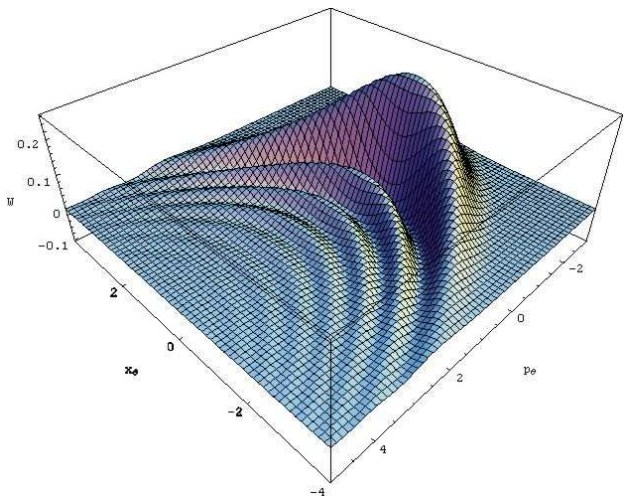}
\end{center}
\caption{$W(x_{\theta},p_{\theta})$, with $r=0.8$, $|\gamma|=0.4$,
$\beta=3$, $\theta=\frac{\pi}{2}$, for the canonical constraint
$\delta-\theta=-\frac{\pi}{2}$,
$\delta+\theta-\phi=\frac{\pi}{2}$.}
\label{W11}
\end{figure}

\begin{figure}
\begin{center}
\includegraphics*[width=8.2cm]{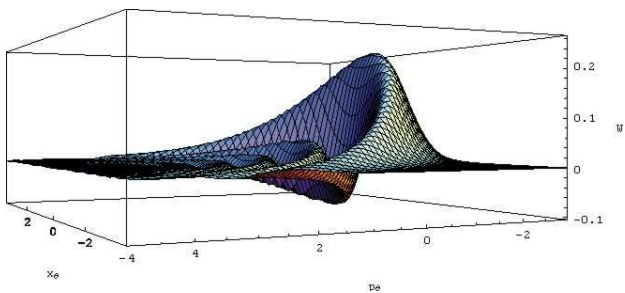}
\end{center}
\caption{$W(x_{\theta},p_{\theta})$, with $r=0.8$, $|\gamma|=0.4$,
$\beta=3$, $\theta=\frac{\pi}{2}$, for the canonical conditions
$\delta-\theta=-\frac{\pi}{2}$,
$\delta+\theta-\phi=\frac{\pi}{2}$.} 
\label{W12} 
\end{figure}

\begin{figure}
\begin{center}
\includegraphics*[width=8.2cm]{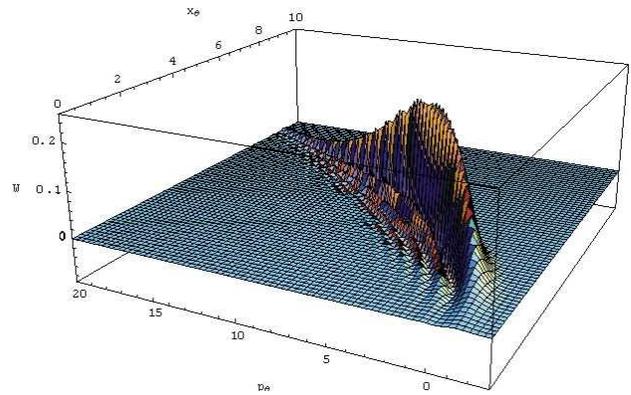}
\end{center}
\caption{Plot of the Wigner function
$W(x_{\theta},p_{\theta})$, with $r=0.8$, $|\gamma|=0.4$,
$\beta=3$, $\theta=\frac{\pi}{3}$, for the canonical
conditions $\delta-\theta=-\frac{\pi}{2}$,
$\delta + \theta - \phi = \frac{\pi}{2}$.}
\label{W2}
\end{figure}

We first consider the quasi--probability distributions
associated to HOMPSS for some values of $\theta$ and compare them
to the corresponding distributions associated to the standard
two--photon squeezed states. This will allow a better understanding 
of the behavior of the photon number distributions and of the 
normalized correlation functions that will be computed later.
The $Q$-function
\be
Q(\alpha)=\frac{1}{\pi} |<\alpha|\Psi>_{\beta}|^{2} \; .
\ee
gives the statistics of heterodyne detection, and
corresponds to a measure of two orthogonal quadrature components
($|\alpha \rangle$ being the coherent state associated
to the coherent amplitude $\alpha = \alpha_{1}
+ i \alpha_{2}$).
Figs.~\ref{Q1} and \ref{Q2} show three--dimensional plots of the
$Q-$function of the HOMPSS, for fixed values of $r$, $|\gamma|$,
$\beta$ and for two different values of $\theta$.
For $\theta=\pi/2$, which corresponds to the case $F=X_{2}^{2}$,
i.e. no mixing, the plot resembles the $Q-$function for a
squeezed state, but we can
observe a deformation of the basis, curved along the
$Re[\alpha] = \alpha_{1}$ axis.
For $\theta=\pi/3$, a case of true mixing of the field
quadratures, the deformation becomes much more evident and
the function is strongly rotated and elongated with
respect to the $Q$-function of the two--photon squeezed state.
Homodyne detection measures a quadrature component
$X_{\lambda}=\frac{1}{\sqrt{2}}(a_{\lambda}+a_{\lambda}^{\dag})$,
where $\lambda$ is a phase determined by the phase of the local
oscillator. Homodyne statistics correspond to project the Wigner
quasiprobability distribution onto a $x_{\lambda}$ axis. With
the identification $\lambda = \theta$ we plot the Wigner
quasi--probability distribution for orthogonal quadrature components
$x_{\theta}$ and $p_{\theta}$:
\begin{equation}
W(x_{\theta},p_{\theta})=\frac{1}{\pi}\int dy e^{-2 i
p_{\theta}y}\Psi_{\beta}^{*}(x_{\theta}-y)\Psi_{\beta}(x_{\theta}+y) \; .
\end{equation}
In Figs.~\ref{W11} and \ref{W12} we show respectively a
global projection and an orthogonal section of the
Wigner function for $\theta=\pi/2$, i.e. no mixing,
fixed canonical constraint, and intermediate
value $|\gamma| = 0.4$ of the nonlinearity.
We see from Figs.~\ref{W11} and \ref{W12} that the Wigner function
displays interference fringes and negative values, exhibiting a
strong nonclassical behavior.
In Fig.~\ref{W2} we show the Wigner
function, for the same values of $r$, $|\gamma|$, $\beta$,
but for $\theta = \pi/3$, a value that realizes a true
mixing of the field quadratures. We see that
the distribution becomes strongly rotated and elongated, with
a pattern of interference fringes and negative values, providing
evidence of the complex statistical structure of the HOMPSS
when a true homodyne mixing is realized.
The behaviors of the quasi--probability distributions suggest the
following considerations. We recall that the HOMPSS,
although being states of minimum uncertainty in the transformed
(``dressed'') modes $(b, b^{\dagger})$, are not minimum
uncertainty states for the original quadratures $X_1$ and $X_2$;
in fact, Eqs. (\ref{dxps})--(\ref{dxps2}) show that a further term
due to the unavoidable statistical correlations adds to the pure
vacuum fluctuations. This fact is reflected in the rotation and in
the deformation of the distributions. But we expect that this
features will affect also the behavior of other statistical
properties, like the photon number distribution. In particular,
shape distortions of the quasi--probability distributions strongly
modify the original ellipse associated to the standard two--photon
squeezed states, giving rise, for suitable values of the
parameters, to deformed intersection areas with the circular
crowns associated in phase space to number states. In turn, this
will lead to modified behaviors of the photon number distribution,
including possible enhanced or subdued oscillations
\cite{schleich} as the mixing angle $\theta$ is varied. Moreover,
we expect that also the second-- and fourth-- order normalized
correlation functions will strongly depend on $\theta$, showing a
deeper nonclassical behavior, for instance antibunching, in
correspondence of values of $\theta$ associated to stronger
nonclassical features of the Wigner function, like negative values
and interference fringes.

\subsection{Photon statistics}

\begin{figure}
\begin{center}
\includegraphics*[width=8.2cm]{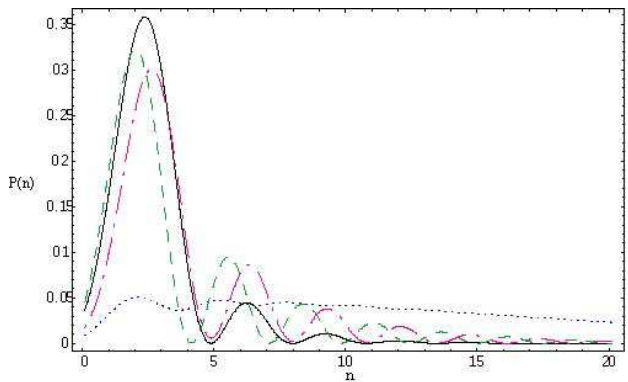}
\end{center}
\caption{$P(n)$ for the HOMPSS, corresponding to the canonical
constraints $\delta-\theta=-\frac{\pi}{2}$,
$\delta + \theta - \phi = -\frac{\pi}{2}$, for
different values of the parameters:
$\beta=3$, squeezing magnitude $r=0.8$, and
strength of the nonlinearity $|\gamma| = 0$
(solid line); $\beta=3$, $r=0.8$, $|\gamma|=0.4$,
$\theta=0$ (dotted line); $\beta=3$, $r=0.8$, $|\gamma|=0.5$,
$\theta=\frac{\pi}{6}$ (dashed line); $\beta=3$, $r=0.5$,
$|\gamma|=0.5$, $\theta=\frac{\pi}{4}$ (dot--dashed line).}
\label{pnd}
\end{figure}

We begin by analyzing the probability for counting $n$ photons, 
the so-called photon number
distribution (PND) $P(n)=|\langle n|\Psi\rangle_{\beta}|^{2}$,
in direct detection, and neglecting detection losses:
\begin{eqnarray}
P(n) & = & \left|\int dx_{\theta} \langle n|x_{\theta}\rangle \langle
x_{\theta} | \Psi\rangle_{\beta}\right|^{2} \nonumber \\
& = & \frac{1}{2^{n} n! \pi^{1/2}}\left|\int dx_{\theta}
e^{-\frac{x_{\theta}^{2}}{2}}H_{n}(x_{\theta})
\Psi_{\beta}(x_{\theta})\right|^{2}.
\end{eqnarray}
Due to the nonlinear nature of the function $F$, it is in general
impossible to write a closed analytic expression for $P(n)$, which
can however be easily determined numerically. In Fig.~\ref{pnd} we
plot the PND of the HOMPSS for various intermediate values of the
local oscillator $\theta$, together with the PND of the standard
two--photon squeezed states.
As foreseen from the previous phase--space analysis, the behavior
of the PND strongly depends on the value of the mixing angle
$\theta$, and for suitable choices of $\theta$, $r$ and
$|\gamma|$, the PND shows larger oscillations with respect to the
standard two--photon squeezed states and the HOMPSS with
$\theta=0$. Moreover, the oscillation peaks persist for growing
$n$ and are shifted for different values of $|\gamma|$; this
behavior is due to the different terms entering in the unitary
operator $U_{\theta}(X_{\theta})$ Eq.~(\ref{Uph3}), which mix in a
peculiar way the quadrature operators for $\theta \neq 0,\pi/2$.

\begin{figure}
\begin{center}
\includegraphics*[width=8.2cm]{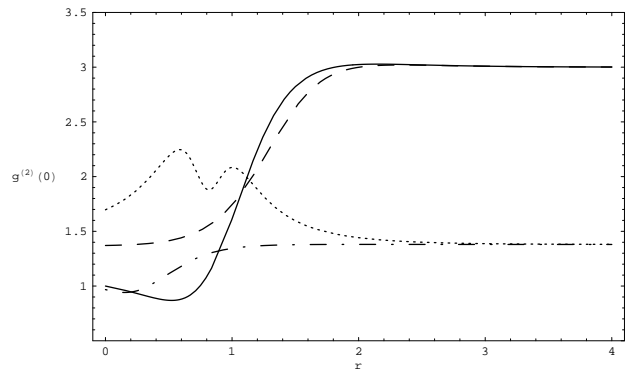}
\end{center}
\caption{The correlation function $g^{(2)}$ as function of $r$ for
the HOMPSS, corresponding to the canonical constraints
$\delta-\theta=-\frac{\pi}{2}$,
$\delta+\theta-\phi=-\frac{\pi}{2}$, for several choices of the
parameters: $\beta=3$, $\gamma=0$ (solid line); $\beta=3$,
$|\gamma|=0.4$, $\theta=0$ (dashed line); $\beta=3$,
$|\gamma|=0.05$, $\theta=\frac{\pi}{6}$ (dot--dashed line);
$\beta=3$, $|\gamma|=0.5$, $\theta=\frac{\pi}{6}$ (dotted line).}
\label{g21}
\end{figure}

\begin{figure}
\begin{center}
\includegraphics*[width=8.2cm]{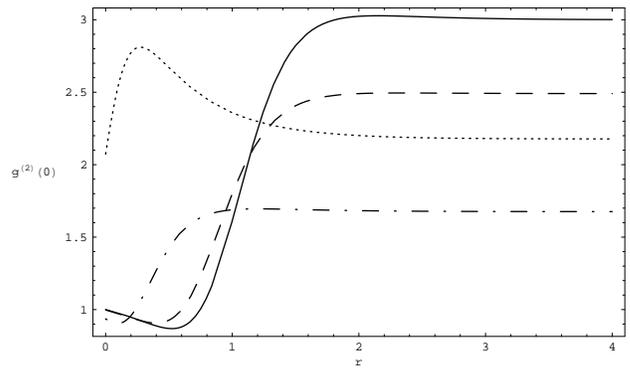}
\end{center}
\caption{The correlation function $g^{(2)}$ as function 
of $r$ for the HOMPSS, corresponding to the canonical conditions
$\delta-\theta=-\frac{\pi}{2}$,
$\delta+\theta-\phi=\frac{\pi}{2}$, versus the two-photon
squeezed states (full line), for different choices of the 
parameters: 1) $\beta=3$,
$|\gamma|=0.05$, $\theta=\frac{4\pi}{9}$ (dashed line); 2)
$\beta=3$, $|\gamma|=0.2$, $\theta=\frac{\pi}{3}$ (dot-dashed
line); 3) $\beta=3$, $|\gamma|=0.5$, $\theta=\frac{\pi}{3}$ 
(dotted line).} 
\label{g22} 
\end{figure}

\begin{figure}
\begin{center}
\includegraphics*[width=8.2cm]{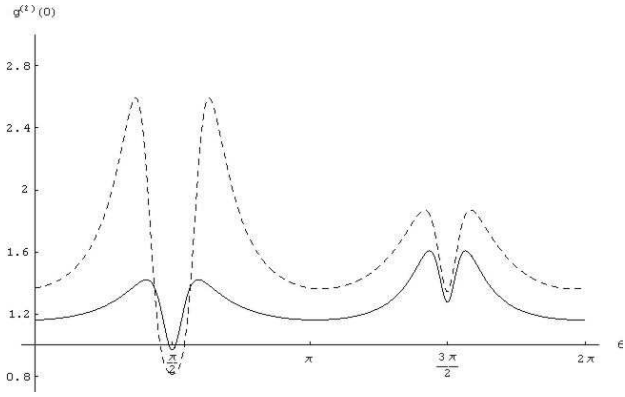}
\end{center}
\caption{The correlation function $g^{(2)}$ as function of $\theta$ 
for the HOMPSS, corresponding to the canonical conditions
$\delta-\theta=-\frac{\pi}{2}$,
$\delta+\theta-\phi=\frac{\pi}{2}$, for: 1) $\beta=3$, $r=0.8$,
$|\gamma|=0.1$ (full line); 2) $\beta=3$, $r=0.5$, $|\gamma|=0.4$
(dashed line).} 
\label{g2th2}
\end{figure}

\begin{figure}
\begin{center}
\includegraphics*[width=8.2cm]{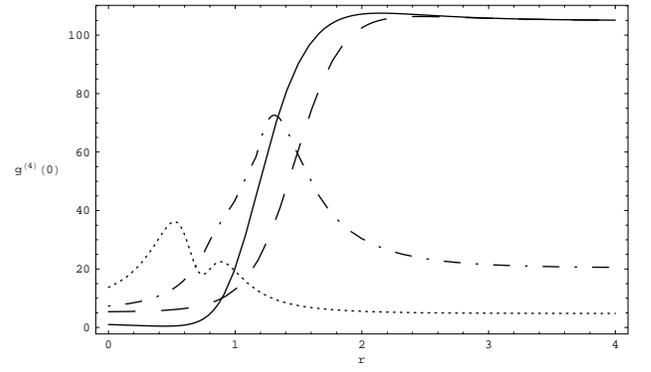}
\end{center}
\caption{The correlation function $g^{(4)}$ as function of $r$ for 
the HOMPSS, corresponding to the canonical conditions
$\delta-\theta=-\frac{\pi}{2}$, $\delta+\theta-\phi=-\frac{\pi}{2}$, 
versus the two-photon squeezed states (full line), 
for several choices of the parameters: 1) $\beta=3$,
$|\gamma|=0.4$, $\theta=0$ (dashed line); 2) $\beta=3$,
$|\gamma|=0.4$, $\theta=\frac{\pi}{12}$ (dot-dashed line); 3)
$\beta=3$, $|\gamma|=0.4$, $\theta=\frac{\pi}{6}$ (dotted line).}
\label{g41}
\end{figure}

\begin{figure}
\begin{center}
\includegraphics*[width=8.2cm]{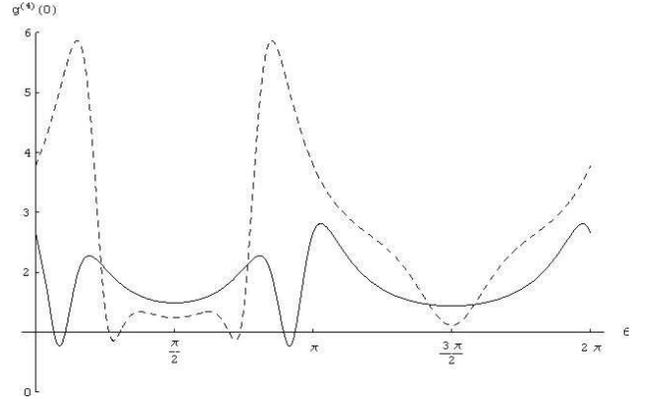}
\end{center}
\caption{The correlation function $g^{(4)}$ as function of 
$\theta$ for the HOMPSS, corresponding to the canonical conditions
$\delta-\theta=-\frac{\pi}{2}$,
$\delta+\theta-\phi=-\frac{\pi}{2}$, for: 1) $\beta=3$, $r=0.5$,
$|\gamma|=0.1$ (full line); 2) $\beta=3$, $r=0.1$, $|\gamma|=0.2$
(dashed line).} 
\label{g4th}
\end{figure}

Concerning the correlation functions, it is interesting to study
the behavior of the normalized second order correlation function
\be 
g^{(2)}(0)=\frac{\langle a^{\dag 2}a^{2}\rangle}{\langle
a^{\dag}a\rangle^{2}}=\frac{\langle a_{\theta}^{\dag
2}a_{\theta}^{2}\rangle}{\langle
a_{\theta}^{\dag}a_{\theta}\rangle^{2}} \; , 
\ee 
and of the normalized fourth order correlation function 
\be
g^{(4)}(0)=\frac{\langle a_{\theta}^{\dag
4}a_{\theta}^{4}\rangle}{\langle
a_{\theta}^{\dag}a_{\theta}\rangle^{4}} \; , 
\ee 
for the HOMPSS, and determine the different physical regimes. 
In Figs.~\ref{g21} and \ref{g22} we compare $g^{(2)}(0)$ as a function 
of the squeezing parameter $r$ for the two--photon squeezed state and for
the HOMPSS at different values of $\theta$.
We see that also the correlation function $g^{(2)}(0)$ shows the
strong nonclassical features of the HOMPSS; both the nonlinearity
strength $|\gamma|$ and angle $\theta$ strongly influence the
behavior of the curves. In fact, the curves deviate from the
standard form and saturate at lower values with respect to TCS.
\begin{figure}
\begin{center}
\includegraphics*[width=8.2cm]{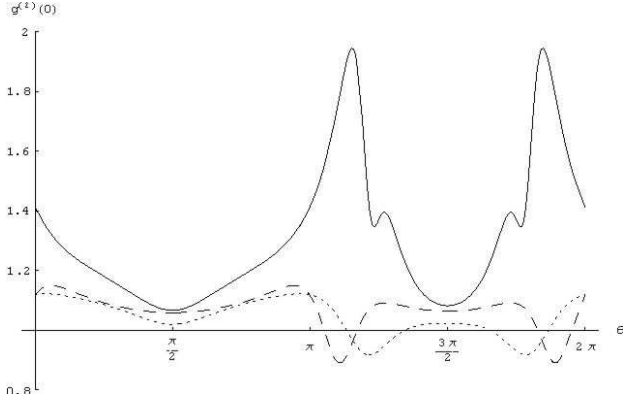}
\end{center}
\caption{The correlation function $g^{(2)}$ as function 
of $\theta$ for the HOMPSS,
corresponding to the canonical conditions
$\delta-\theta=\frac{\pi}{2}$, $\delta+\theta-\phi=\frac{\pi}{2}$,
for: 1) $\beta=3$, $r=0.5$, $|\gamma|=0.4$ (full line); 2)
$\beta=3$, $r=0.4$, $|\gamma|=0.1$ (dashed line); 3) $\beta=3$,
$r=0.1$, $|\gamma|=0.1$ (dotted line).} 
\label{g2th1} 
\end{figure}
The most significant feature is however obtained  plotting
$g^{(2)}(0)$ as a function of $\theta$, Fig. \ref{g2th1}, Fig.
\ref{g2th2}; the plots clearly demonstrate that it is possible to
pass from a subpoissonian to superpoissonian statistics. So,
tuning $\theta$, we can have photon bunching or antibunching and
we can select the preferred statistics.
Similar considerations can be made about the fourth order
correlation function $g^{4}(0)$, as it can be
seen by Fig.~\ref{g41}, where it is shown the dependence
from the squeezing parameter $r$, and
by Fig.~\ref{g4th}, where instead $g^{4}(0)$
is plotted as a function of the angle
$\theta$. Also the normalized fourth--order correlation
function shows a particular shape for intermediate value of
$\theta$ and varying saturation levels, depending on the
parameters of the nonlinear term. Moreover, we have four photon
bunching or anti-bunching in correspondence of
different values of $\theta$.

\section{Conclusions and outlook}

\label{outlook}

We have introduced single-mode nonlinear canonical transformations,
which represent a general and simple extension of the linear Bogoliubov
transformations. They are realized by
adding a largely arbitrary nonlinear function of the
homodyne quadratures $X_{\theta}, P_{\theta}$. We have introduced
the Homodyne Multiphoton Squeezed States (HOMPSS) defined as the
eigenstates of the transformed annihilation operator. The HOMPSS
are in general non Gaussian, higly nonclassical states which
retain many properties of the standard coherent and squeezed states;
in particular, they constitute an overcomplete basis in Hilbert
space. On the other hand, many of their statistical properties can
differ crucially from the ones of the Gaussian states. In particular,
we have shown the strong dependence of the photon statistics on the
local oscillator angle $\theta$.
Among other remarkable features, there are the possibilities of
exploiting the homodyne angle as a tuner to select
sub--poissonian or super--poissonian statistics, and as catalyzer
enhancing the average photon number in a state.

The single--mode multiphoton canonical formalism selects a large number
of non Gaussian, nonclassical states, including the single--mode cubic
phase state, which generalize the degenerate, Gaussian squeezed states.
On the other hand, both from the point of view of modern applications, 
as e.g. quantum computation, and for experimental implementations,
generalizations to two and many modes are of great interest.

In the following companion paper, "Structure of multiphoton
quantum optics. II. Bipartite systems, physical processes, and
heterodyne squeezed states" (Part II), we will extend the canonical
scheme developed in the present paper (Part I) to study multiphoton processes
and multiphoton squeezed states for systems of two correlated modes of the
electromagnetic field. This extension is important and desirable in view of 
the modern developments in the theory of quantum entanglement and quantum 
information. In particular, we will show how to define two-mode nonlinear 
canonical transformations and we will determine the associated 
``heterodyne multiphoton squeezed states'' (HEMPSS).
In the context of macroscopic (quantum) electrodynamics in nonlinear media,
we will moreover discuss the kinds of multiphoton processes that can allow 
the  experimental realizability of the HEMPSS and of the effective 
interactions associated to the two--mode nonlinear canonical formalism.

\end{document}